# Spike Camera and Its Coding Methods

Siwei Dong, Tiejun Huang and Yonghong Tian

*EECS, Peking University*
*No.5 Yiheyuan Road, Haidian District*
*Beijing, 100871, P.R. China*
*{swdong, tjhuang, yhtian}@pku.edu.cn*

*Abstract*: This paper introduces a spike camera with a distinct video capture scheme and proposes two methods of decoding the spike stream for texture reconstruction. The spike camera captures light and accumulates the converted luminance intensity at each pixel. A spike is fired when the accumulated intensity exceeds the dispatch threshold. The spike stream generated by the camera indicates the luminance variation. Analyzing the patterns of the spike stream makes it possible to reconstruct the picture of any moment which enables the playback of high speed movement.

## 1. Introduction

One of the perceptions of the world for human and animals is depended on their eyes to capture light. Modern cameras use CCD (charge-coupled device) or CMOS (complementary metal oxide semiconductor) to capture light and record the motion, which results in a large number of digital pictures and videos [1]. Conventional motion pictures are expressed as two-dimensional images and video sequences. Digital cameras were invented to simulate human eyes with the ability to capture still and moving pictures. However, the cameras cannot record every moment instantly which leads to discrete simulation as a compromise. As the human visual system can process 10 to 12 separate images per second and perceive them individually, and sequences at higher rates are perceived as motion [2], the common frame rates or frame frequencies are 25, 30, 60, 120, etc. Although there are high speed cameras which are capable of image exposures in excess of 1/1,000 second or frame rates in excess of 250 frames per second [3], the motion changes of high speed objects in microseconds are missing. For scenarios of surveillance of traffic accident and suspect identification, every moment is important, but the key frame may be blurred due to limited frame rate of the capture system.

There several ways to solve the problem discussed above. One is to increase the frame rate of the camera which is very difficult because of the limit associated with the time cost of the image exposure and the accuracy of the current measurement [4]. Larger frame rate means higher shutter speed and more noise. Another way is developing an asynchronous vision sensor that the output at each pixel is asynchronous which is in the form of address-events. The architecture is inspired from the emerging neural coding in the retina. The ganglion cells of retina receive the information from the photoreceptors and respond with a short burst of spikes which play a role in the propagation of the signals to the primary visual cortex and higher brain regions [5][6]. Although the conventional camera systems are doing the basic simulation of retina, utilizing the spikes to visual coding and texture reconstruction enlarges the horizon for high speed imaging and vision processing.

We design a spike camera which has an array of sensors, each of which captures one pixel independently and keeps recoding the luminance intensity. If the accumulated intensity reaches the dispatch threshold, a spike is dispatched to the bus. Different from conventional digital cameras, the spike camera does not care about the quality of image exposure but focuses on the luminance intensity variation, therefore the temporal resolution can be as high as possible.

The rest of the paper is organized as follows: Section 2 briefly reviews related work, and Section 3 presents the design of the spike camera including two methods of texture decoding. In Section 4, the results of the simulation experiment are demonstrated. Finally, the paper is concluded in Section 5.

## 2. Related Works

### 2.1. Dynamic Vision Sensor

The video data captured from conventional cameras in our daily life is frame-based architecture which consists of a sequence of pictures [7]. With the rapid development of integrated circuit technology, the frame rate of cameras becomes much higher which as a consequence, directly leads to the emergence of the high speed cameras. Although the frame-based cameras are faster than ever before, they are still unlike the human's eyes or retinal system. Recently, Dr. Delbruck's team has developed a bio-inspired dynamic vision sensor (DVS) camera named DAVIS [8][9] which simulates the retina and generates spikes asynchronously when the luminance changes at each pixel. The DAVIS camera is able to reduce the temporal redundancy and the spikes are quite sparse that are in the form of the address-events representation (AER, the events are encoding with the x- and y-addresses of the pixel) [10]. Each pixel can fire two kinds of events: on-event if the luminance intensity increases and off-event if the intensity decreases. The intensity at the pixel is reset when an event is fired. Figure 1 demonstrates the on/off-event of object movement (white points for on-event while black ones for off-event). The DAVIS camera is natively suitable for motion-based tasks such as action detection and gesture recognition. Meanwhile, the AER protocol makes it capable of high-speed capturing and asynchronous transmission.

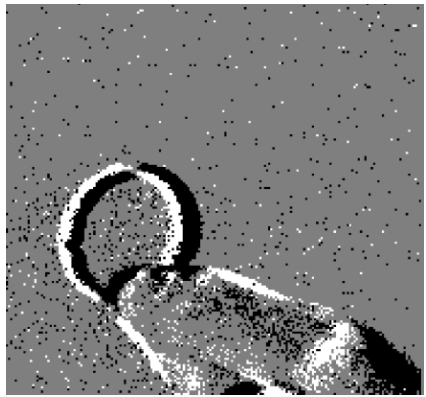

Figure 1: a ring moves from left side (captured via DAVIS 240B)

Different from Dr. Delbruck's work, Prof. Chen's group from Nanyang Technological University develops another DVS camera named CeleX with higher resolution of 384x320 [11]. The spike generated from their camera not only carries the address and the timestamp, but also brings a nine-bit ADC (analog-to-digital converter) value. There's no off-event in their design, thus the spike itself only indicates the luminance intensity change, but the attached ADC value makes up the missing of off-event. As a consequence, the CeleX camera is not only able to demonstrate the motion change of objects (shown in Fig. 2a), but also capable of texture reconstruction (shown in Fig. 2b).

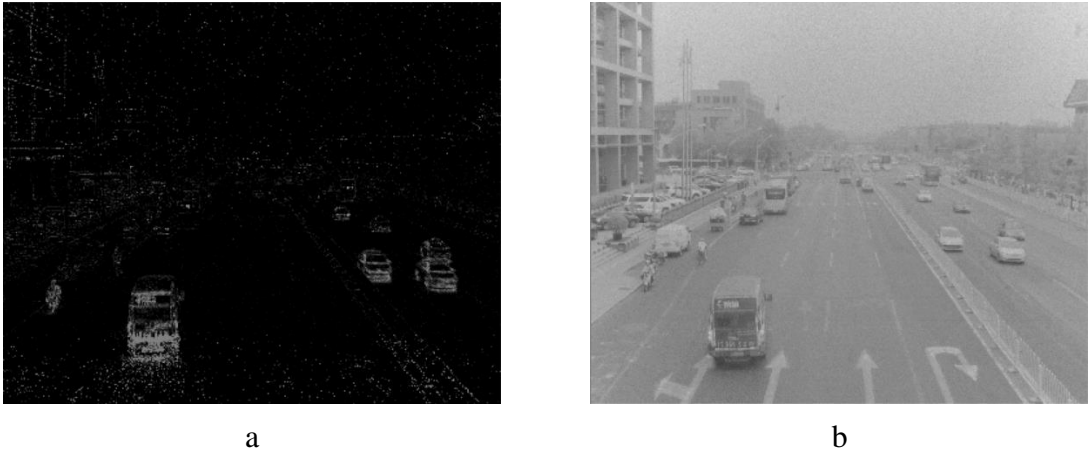

          a                    b

Figure 2: the Zhongguancun North Street (captured via CeleX)

## 2.2. Processing of spikes

DVS cameras encode the temporal contrast of luminance intensity into address-events which are asynchronously transmitted for subsequent processing. The row and column addresses, the timestamp, as well as the polarity (on/off) are three main components of the output event data [12]. If we collect all the spikes/events fired at the same time and draw points according to the x- and y-addresses of the spikes on a clear screen, as the time goes by, the scene structure and the object movement can be illustrated. However, the texture is missing due to the spike itself only indicating the luminance change, if no ADC data is transmitted along with the spike. The spikes are often used to track moving objects and to detect specific actions by analyzing the accurate timing information preserved in the spikes. Although it is quite useful in motion applications, the DVS camera faces big challenges in processing still objects. If there is no object moving, the luminance intensity does not change much, leading to no spike to fire. In this case, the conventional frame-based cameras are better at processing still scenario while the DVS cameras are up to motion-related tasks. This paper explores a novel route to take advantage of the DVS sensor but as well to reconstruct visual-friendly pictures.

## 3. The Design of Spike Camera and Texture Decoding

The spike camera is event-based of which the pixels are independent. Once the analog-to-digital converter (ADC) completes the signal conversion and outputs the digital luminance intensity, the accumulator at each pixel accumulates the intensity. For a pixel, if the accumulated intensity reaches the dispatch threshold $\Omega$, a spike is fired indicates that the luminance change here is large enough. At the same time, the corresponding accumulator is reset which subtracts the value of the dispatch threshold from its own intensity. Eq. 1 is the model of the accumulator, where $A_{t_i}$ and $A_{t_{i-1}}$ refer to the values of the accumulator at the moment of $t_i$ and $t_{i-1}$ respectively, and $I_{t_i}$ is the input ADC value at $t_i$.

$$A_{t_i} = (A_{t_{i-1}} + I_{t_i}) \bmod \Omega \tag{1}$$

For example, shown in Fig. 3, we assume that the dispatch threshold $\Omega = 12$, and at each moment the accumulator gets about 5 units of intensity increased. After about 3 moments (at the end of $t_3$), the accumulated intensity is 14 units, then a spike is dispatched and the intensity subtracting the dispatch threshold $\Omega$ becomes 2 units. Then at the end of $t_4$, as 5 units is newly collected, the accumulated intensity is 7 units. Thus, the dispatched spikes are 0, 1, 0, 1, 1, 0, 0, 1, 0 ... Here "0" means no spike dispatched and "1" refers to spike. The spike stream can be written as in Eq. 2.

$$S_{t_i} = \begin{cases} 1, & \text{if } A_{t_{i-1}} + I_{t_i} \geq \Omega \\ 0, & \text{otherwise} \end{cases} \tag{2}$$

Compared to the conventional camera, the spike camera only cares about the change of luminance intensity. Considering that at different pixel, the accumulating speed of the luminance intensity is quite different; the patterns of dispatched spike stream are as well different from each other. For brighter pixels, "1" appears more frequently than that of darker ones. The idea is easy to explain. The brighter pixel indicates that more photons are collected at the pixel resulting in larger ADC value which is easier and faster to exceed the dispatch threshold. According to this principle, the texture can be decoded by analyzing the patterns of spikes. It is quite similar to the processing of the responses of ganglion cell which illustrates the outline of the object by decoding the spike latencies [6].

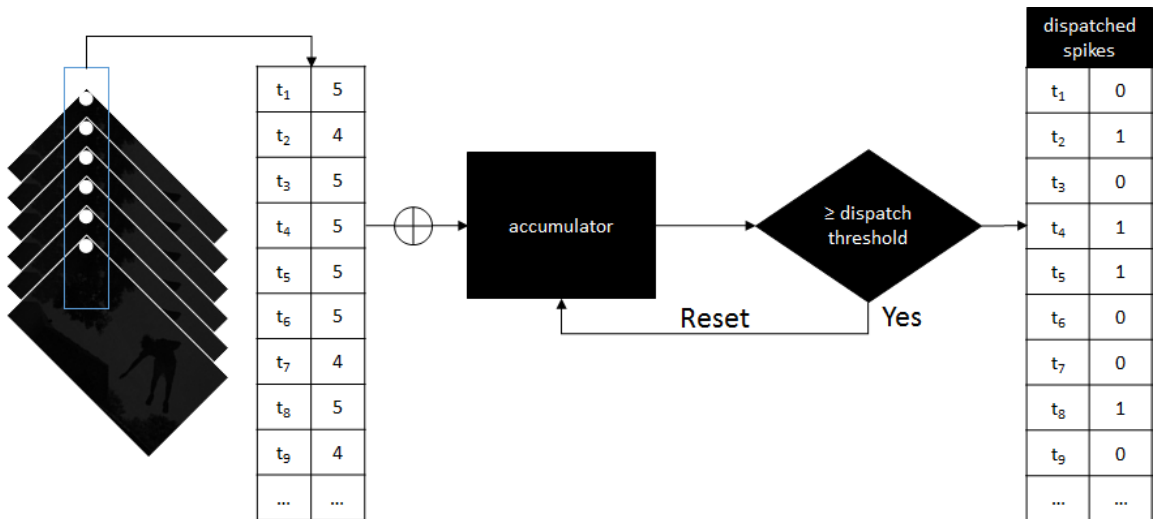

Figure 3: The working flow of the spike camera

To decode the texture of any historical moment, there are two kinds of ways for different application purposes. For real-time applications, the texture is decoded from the latencies of spikes, shown in Fig. 4.

$$P_{t_i} = \frac{\Omega}{d_{t_i}} \qquad (3)$$

$P_{t_i}$ refers to the texture of the pixel at the moment of $t_i$, and $d_{t_i}$ means the latency between $t_i$ and the last moment when spike appears. Compared to the original captured texture data, the decoded values are not perfectly matched. But the variation is in accordance with the original one. This method (texture from latencies, TFL) can reconstruct the outline of the texture but not the clear details. When the object moves very quickly, the picture reconstructed from latencies performs the motion nearly synchronously.

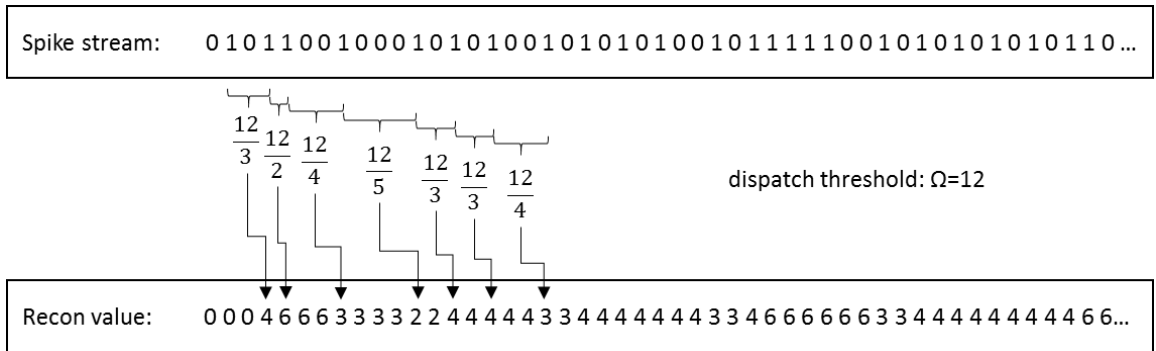

Figure 4: the texture decoded from spike latencies (TFL)

As spikes are dispatched with a very high frequency, if we play back the spikes, the historical pictures are able to be illustrated. In this method (texture from playback, TFP), there's a moving time window collecting the spikes in a specific period. By counting these spikes, the texture is computed as:

$$P_{t_i} = \frac{N_w}{w} \cdot C \qquad (4)$$

The size of the time window is $w$ which refers to the previous $w$ moments before $t_i$. $N_w$ is the total number of spikes collected in the time window. $C$ refers to the maximum contrast level which is normally set to 256 (8 bits). The larger size means the playback period is longer and more spikes are included. When the time window size is set to the dispatch threshold $\Omega$, the textures are accurately reconstructed. Meanwhile, the TFP method could reconstruct the texture with various dynamic ranges by resizing the time window to the value of different contrast levels [13] [14].

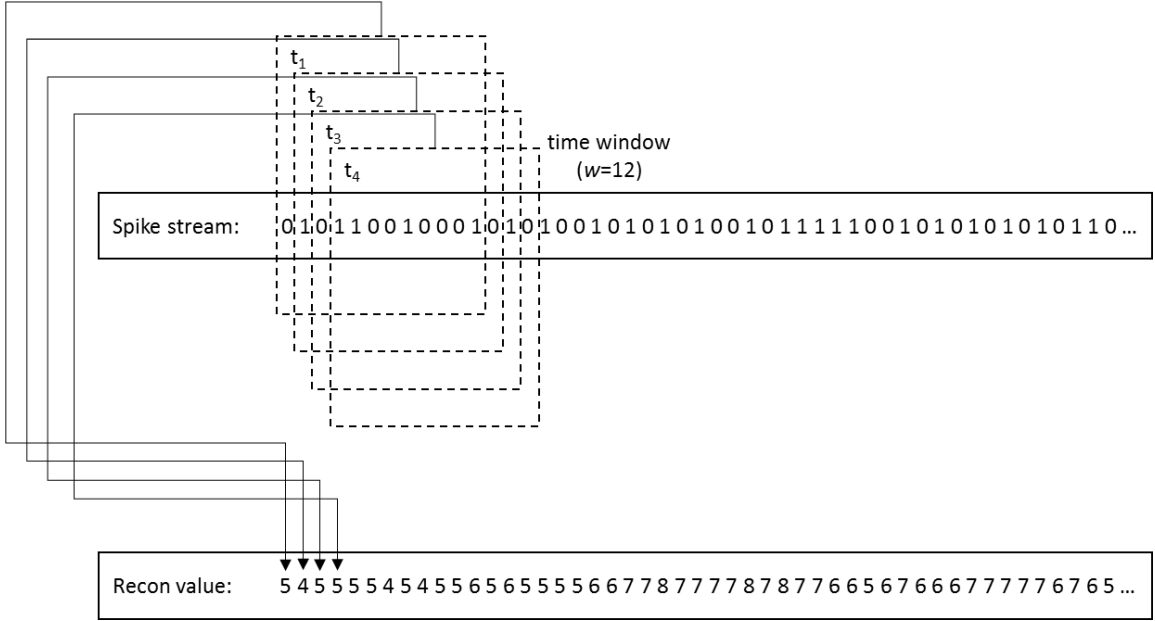

Figure 5: the texture decoded from playback with the moving time window (TFP)

## 4. Experimental Results

### 4.1. Experimental setup

Table 1: the input high speed videos

| Sequences | Frame rate (fps) | Resolution | Snapshot |
|---|---|---|---|
| Somersault | 20000 | 1920×1080 | |
| Milk | 1000 | 1920×1080 | |
| Moto | 3000 | 1280×720 | |
| Waterdrop | 2000 | 320×240 | |

In order to test the performance of the spike camera, especially texture reconstruction, we use several high speed video sequences as the input data, shown in Table 1. The pictures are fed to the spike camera which has an array of sensors with the same size of the video resolution. The dispatch threshold $\Omega$ is set to 256. Each sensor processes the corresponding pixel of the pictures in temporal. As a result, the output spikes of each sensor are combined to form the spike stream. For a specific moment, if no spike is dispatched, "0" will be filled into the stream.

## 4.2. The compression performance

The single spike stream of each sensor is periodic, so it can be effectively compressed. Table 2 demonstrates that the compression ratios of the spike stream via the classic LZ77 (Lempel-Ziv compression algorithm) [15] and LZMA (Lempel-Ziv-Markov chain algorithm) [16] lossless encoders.

Table 2: the compression performance of the spike streams

| Sequences | Spike stream size | Compressed via LZ77 | Compressed via LZMA |
|---|---|---|---|
| Somersault | 2565.9MB | 314.9MB | 264.4MB |
| Milk | 759.4MB | 110.5MB | 93.7MB |
| Moto | 130.5MB | 30.6MB | 24.1MB |
| Waterdrop | 44.9MB | 10.0MB | 8.2MB |

## 4.3. Texture decoding

One of the most important applications of the spike camera is for visual-friendly viewing. The texture reconstruction experiment is conducted utilizing two methods: TFL and TFP. In the TFL experiment, only luminance component (grayscale) is reconstructed, shown in Fig. 6. The result indicates that the outline of the object is quite clear, while some detailed textures are missing. Thus, TFL is more suitable for real-time applications and some detection related tasks.

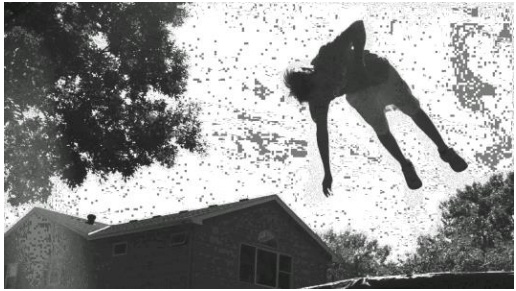
Somersault

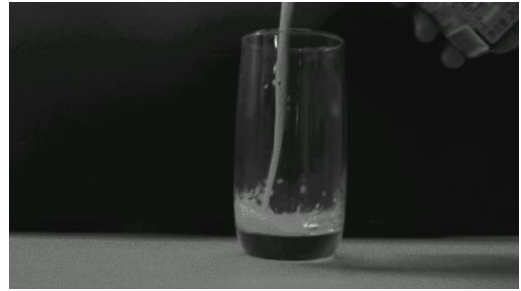
Milk

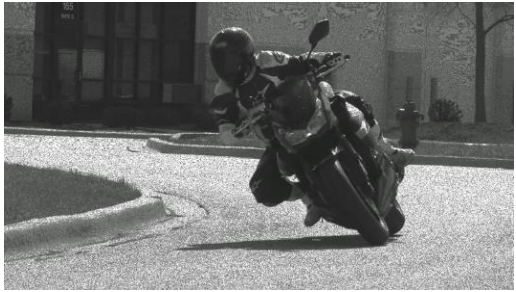
Moto

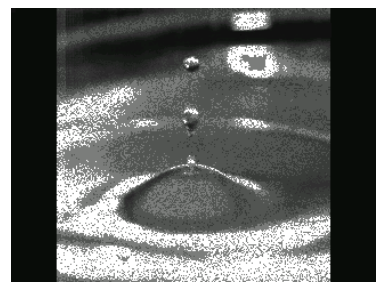
Waterdrop

Figure 6: Texture decoding via TFL

For TFP method, the size of time window is a key parameter in texture reconstruction. We select four typical sizes for comparison which are 32, 64, 128 and 256. From Fig. 7, smaller time window size achieves more clear outline for moving objects but lower quality for static background pixels.

| Time window | Somersault | Milk | Moto | Waterdop |
|---|---|---|---|---|
| 32 | 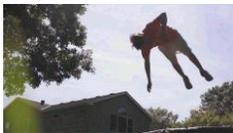 | 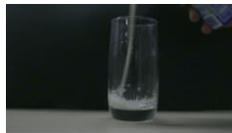 | 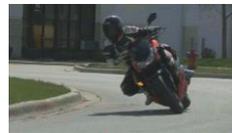 | 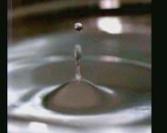 |
| 64 | 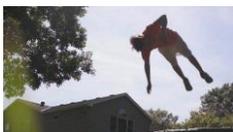 | 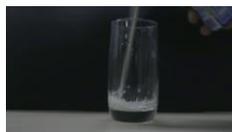 | 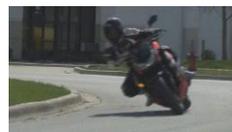 | 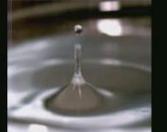 |
| 128 | 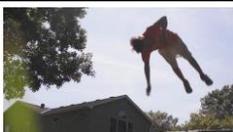 | 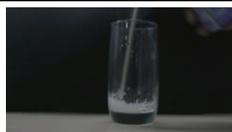 | 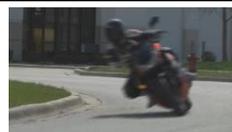 | 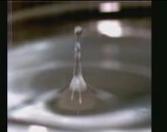 |
| 256 | 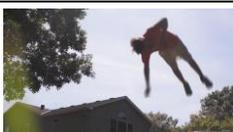 | 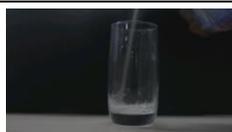 | 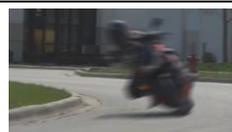 | 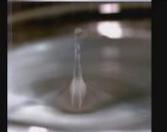 |

Figure 7: texture decoding via TFP

## 4.4. Simulation of dynamic range

In the section, we use an under-exposure video (Fig. 8) as the input for the spike camera. Using TFP method, the visual-friendly picture sequence can be reconstructed and by adjusting the playback time window, the reconstruction performances of different dynamic ranges are tested. Considering that the input video sequence is under exposed, the dispatch threshold is set to a relatively small value of 32. For spike stream decoding, the playback time window is set to different sizes $w = \{32, 64, 128, 256, 512\}$. The reconstruction results in Fig. 9 show that the spike stream coded with $\Omega = 32$ can be decoded into texture pictures with higher dynamic ranges than that of the original under-exposure video. From the results, the static region is reconstructed with more details when higher dynamic range is gained. The motion of the object is blurred in $w = 512$ due to that the exposure period (equals to $w$) is much longer than that of the original under-exposure pictures.

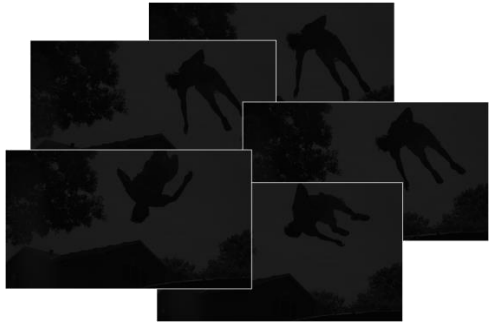

Figure 8: the under-exposure video sequence

| Time window | Dynamic range | Texture decoding |
|---|---|---|
| 32 | [0, 31] | 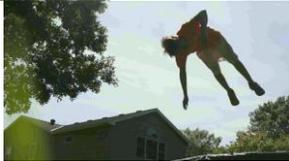 |
| 64 | [0, 63] | 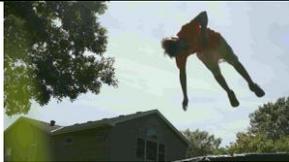 |
| 128 | [0, 127] | 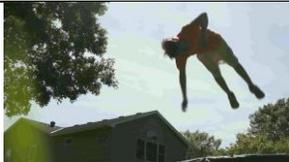 |
| 256 | [0, 255] | 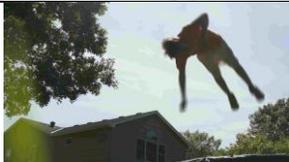 |
| 512 | [0, 511] | 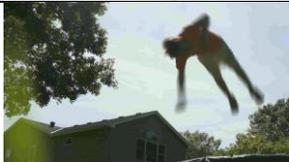 |

Figure 9: texture decoding from the spike stream generated from under-exposure video

## 5. Conclusion

In this paper, a novel event-based spike camera is introduced which simulates the retinal imaging. Two decoding methods of spike stream for texture reconstruction are proposed which enables playing back any historical moment. Experimental results show that TFL is more suitable for real-time applications and object detection or action recognition related tasks, while TFP is capable of reconstructing visual-friendly picture sequences for viewing. For future work, more efficient compression methods especially lossy compressions need to be explored. Moreover, how to enhance the quality of reconstructed texture also need to be studied.